# An express monitoring procedure for low pressure MWPC efficiency value in heavy ion induced complete fusion nuclear reactions

Yu.S.Tsyganov[a], D.Ibadullayev[a,b,c,*], A.N. Polyakov[a], V.B.Zlokazov[a]

[a] Joint Institute for Nuclear Research, 141980 Dubna, Russian Federation.

[b] Institute of Nuclear Physics, 050032 Almaty, Kazakhstan.

[c] L.N. Gumilyov Eurasian National University, 010000 Astana, Kazakhstan.

*- Corresponding author: post address: 141980 Joliot-Curie 6, Dubna, Moscow Region, Russia;

e-mail: tyra@jinr.ru

**Abstract**

A simple routine is proposed for monitoring the efficiency value of a low-pressure pentane-filled multi-wire proportional chamber (MWPC) in long-term experiments. The proposed algorithm utilizes a two-parameter approximation for the background function. It is based on a linear approximation of the background in the energy range of 4.8 to 10.0 MeV and an exponential approximation of the neutron-induced tail in the 1.5 to 4.2 MeV region. This specific energy interval is used to measure the efficiency value. Prior to discussing the algorithm, a description of the DGFRS-2 setup detection module is provided. Additionally, an example of its application is presented for the complete fusion of a heavy-ion-induced $^{232}$Th+$^{48}$Ca → Ds* complete fusion nuclear reaction. Descriptions of two other two-parameter functional dependencies for background approximations are also provided. A feature of this algorithm is that there is no interruption of the main experiment for a calibration test (reaction). An alternative scenario is considered in brief too. In consist of a measurement of scattered target like ions in ~ 30 to 45 MeV energy interval to estimate an efficiency value.

## 1. Introduction

Experiments utilizing gas-filled electromagnetic separators and corresponding detection systems have successfully synthesized elements with atomic numbers 113 to 118. These experiments involved the use of $^{48}$Ca ions [1-4]. By employing improved detection systems, rare alpha and spontaneous fission decays of superheavy nuclei (SHN) were isolated from background events in reactions such as $^{48}$Ca + actinide target → SHN + xn, which were carried out at the U-400 cyclotron, FLNR JINR [4-5]. The cross sections for these reactions varied from 0.1 to 10 picobarns. However, it is anticipated that the production cross-sections of SHN will significantly decrease in future experiments utilizing $^{50}$Ti and $^{54}$Cr ion beams. This necessitates stricter requirements for the properties of the separator and detection system. Consequently, the study of rare decay events of superheavy nuclei and the investigation of their detection characteristics are becoming increasingly important. Moreover, the method of active correlations [4, 6-9], used to suppress background, becomes particularly vital when intense beams of heavy ions (up to 5-10 pµA) are employed.

## 2. Detection module and algorithm for MWPC efficiency estimation

During long-term experiments at the DGFRS-1,2 setups [1,10], the typical time interval between two calibration processes may equal one to two months or even more, thanks to the high stability of the DSSD detector and the entire detection system [11]. The reaction $^{nat}Yb+^{48}Ca \rightarrow$ $Th^*$ is usually used due to its high yield of Th recoils. On the other hand, for the reactions of interest, where we monitor the MWPC detection efficiency value, we use energy intervals of 1.5-4.2 MeV (or sometimes 2-4.2 MeV). This choice is based on the fact that this energy interval corresponds to a higher statistical count of registered events. In Figure 1, the schematic of the DGFRS-2 detection module is shown.

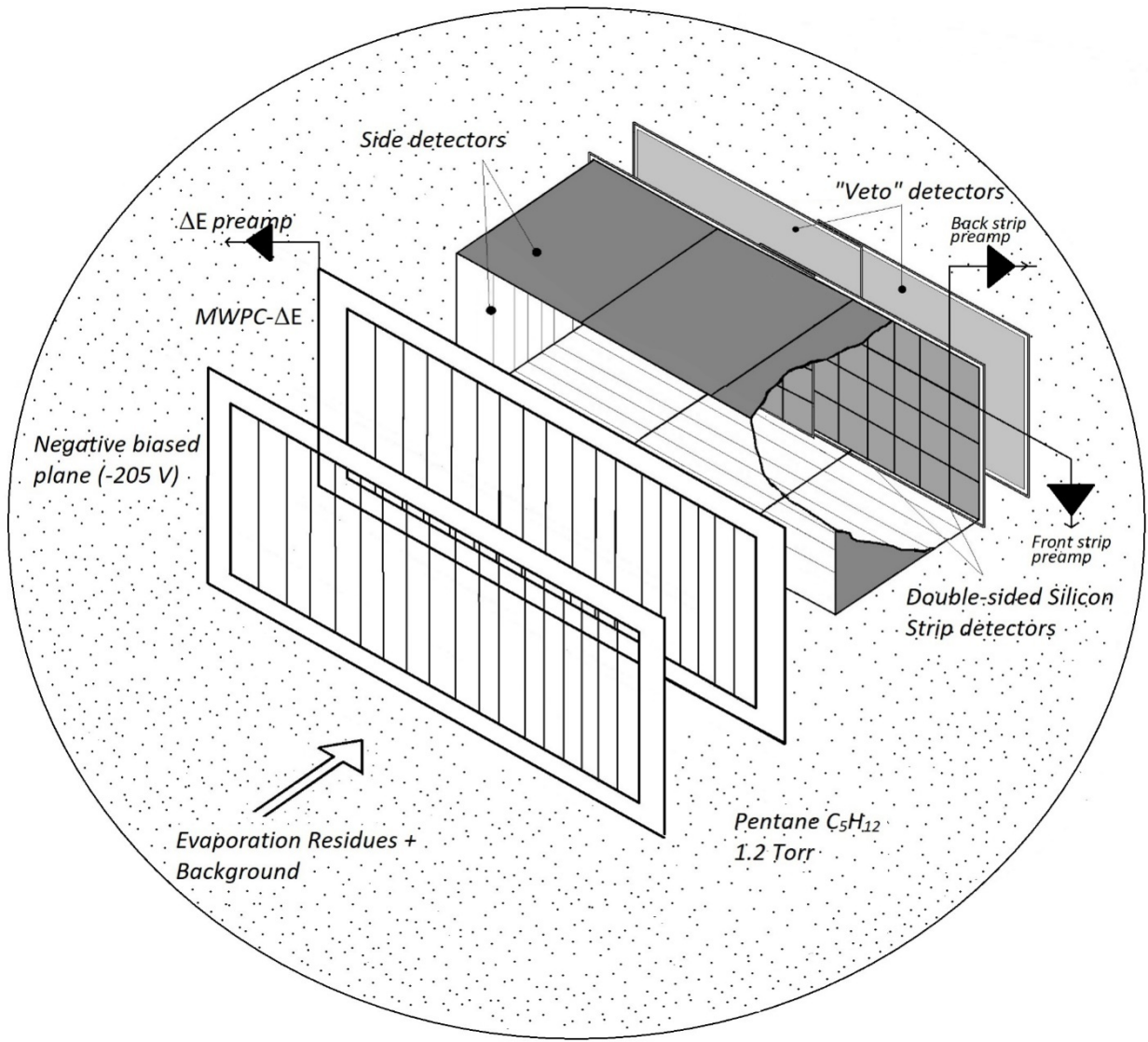

Figure 1. Schematics of the DGFRS-2 detection module (see in details [10-11] )

The detection module consists of a DSSD focal plane with a 48x128 strips detector and a low-pressure MWPC filled with pentane. Its purpose is to detect evaporation residues and their α-decays [8, 11-13]. Surrounding the focal plane are eight Si detectors, each 500 μm thick, with an active area of 60 × 120 mm$^2$. It is important to note that a detailed description of the entire detection system can be found in [11]. Figure 2 depicts typical approximation images for both the neutron-induced region and the α-particle region. To minimize errors associated with simple linear approximations in the ~4.8-9 MeV range, an interactive recalibration process is employed using the VMRIA-2019 code. Using this codeprovidesaprecisedeterminationof α-peakcontents[14-15].

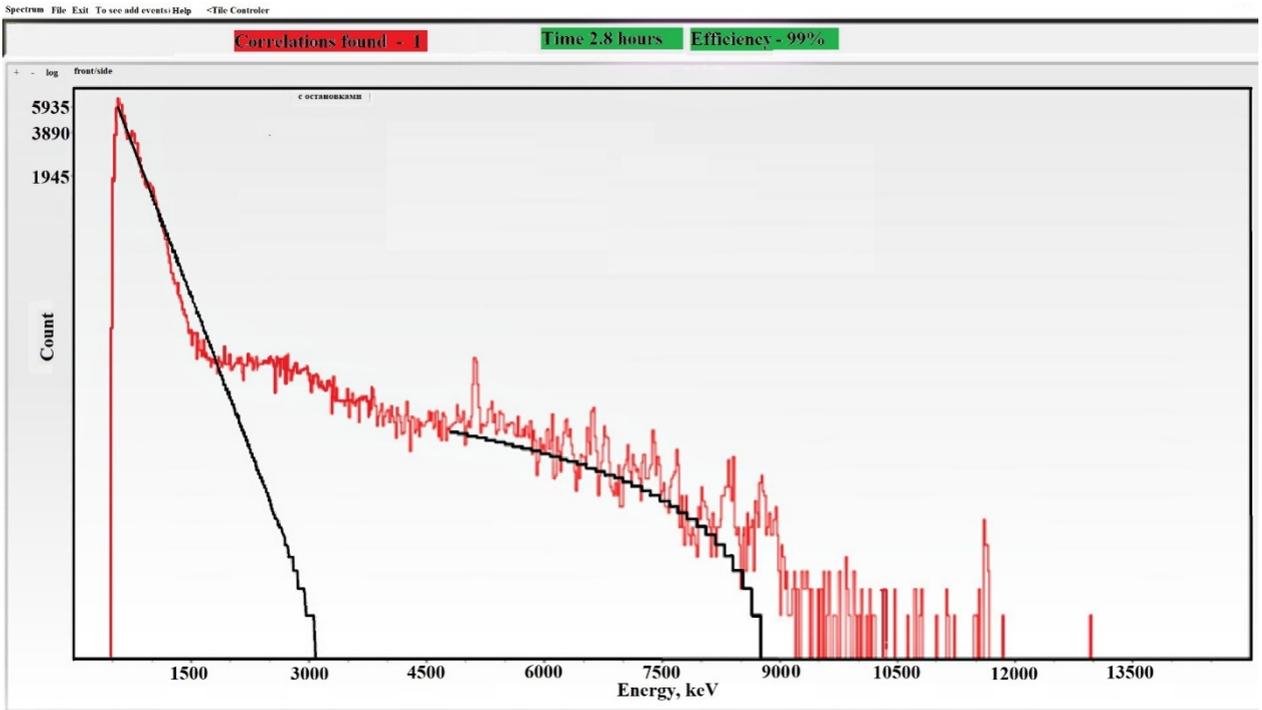

Figure 2. Typical approximation lines for backgrounds. The estimated efficiency parameter is of ~99% shown in upper window. Reaction $^{232}$Th+$^{48}$Ca→Ds*

In Table 1, the relative deviation parameters are presented for the recalibrating procedure. Figure 3 shows the results of applying the VMRIA code [14-15] to one file (No.734) from the $^{232}$Th+$^{48}$Ca → Ds* complete fusion nuclear reaction. The mean value of ξ, which is 0.77, was inserted into the execution code after the recalibration.The value of ξ, therefore, means the ratio of the determined number of α-decays to a similar value calculated using the VMRIA program (Formula 1).

$$\xi = \frac{\sum_{i \to 4.9}^{i \to 9}(n_i - bckg_i)}{\sum_{j=1}^{N} peak_j}. \qquad (1)$$

Here, $n_i$ – content in channel $i$,

$bckg_i$ – background in channel $i$ for linear presentation,

$peak_j$ – content in peak $j$ according to VMRIA code,

$N$ – number of peaks find by VMRIA code.

**Table 1. Recalibration parameter for α-region 4.9 – 9 MeV**

| ξ, function / File No. | 725 | 734 | 786 | 787 | 793 | Mean/std |
|---|---|---|---|---|---|---|
| A·x + B | 0.896 | 0.845 | 0.590 | 0.832 | 0.688 | 0.77/0.13 |
| A·exp(-B·x) | 1.21 | 1.39 | 1.59 | 1.93 | 1.77 | 1.59/0.29 |
| A/$x^2$ + B | 1.19 | 1.27 | 1.28 | 1.60 | 1.60 | 1.39/0.20 |
| A/$x^{1.45}$ + B | 1.15 | 1.22 | 1.21 | 1.52 | 1.52 | 1.32/0.18 |
| A/$x^{1.20}$ + B | 1.12 | 1.16 | 1.14 | 1.43 | 1.43 | 1.26/0.16 |
| A/$x^{0.5}$ + B | 1.01 | 1.01 | 0.94 | 1.18 | 1.01 | 1.03/0.09 |
| $\chi^2$ (in VMRIA representation) | 1.68 | 1.62 | 1.82 | 1.41 | 1.65 | - |
| Intensity $^{48}$Ca+$^{10}$, pμA | 3.2 | 3 | 3 | 3 | 3.2 | 3.08±0.11 |

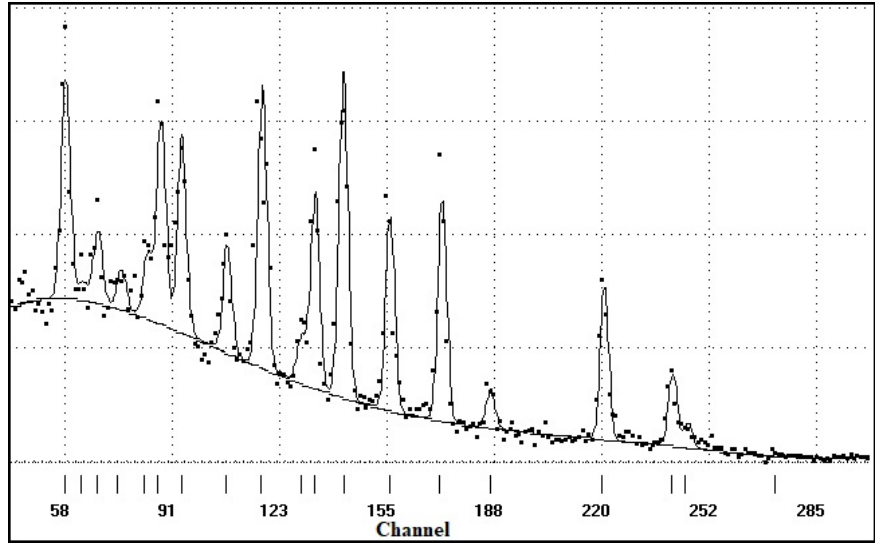

Figure 3. Typical processed spectrum with VMRIA code (*File 734 Tab.1*).

It should be noted that for each approximation function listed in the table, the conditions for searching for both parameters A and B are as follows:

$$\begin{cases} \overline{Y_{01}} = f(X_{01}) \\ \overline{Y_{02}} = f(X_{02}) \end{cases} \tag{2}$$

Here: $X_{01}$=4.8 MeV, $X_{02}$=9 MeV.

And:

$$\begin{cases} \overline{Y_{01}} = 0.125 \cdot \sum_{i=-4}^{3} f(X_i), & X_i = X_{01} + 20 \cdot i \\ \overline{Y_{02}} = 0.125 \cdot \sum_{j=-4}^{3} f(X_j), & X_j = X_{02} + 20 \cdot j \end{cases} \tag{3}$$

Regarding the parameterization of the low-energy tail of the signal distribution above the registration threshold (≥0.5 MeV), which we attribute to neutron-induced events, we apply an exponential distribution (A·exp(-B·x)), as shown by the dotted line on the left side of Figure 2). In this case, the parameters A and B are determined by averaging over four pairs of points, with a step of 100 keV. Specifically, starting from 600 keV, we form four pairs of equations as follows:

$$A \cdot \exp(-B \cdot x_i) = Y_i \text{ and } A \cdot \exp(-B \cdot (x_i + step)) = Y_{i+1}, i = 1 \dots 4. \qquad (4)$$

And, respectively,

$$\bar{A} = 0.25 \cdot \sum A_i, \bar{B} = 0.25 \cdot \sum B_i. \text{ Here, step=100 KeV.} \qquad (5)$$

Note that the relationship between the measured number of α-particles and the number of escaping α-particle signals in the 1.5-4.2 MeV interval was established through the analysis of a decaying off-line spectrum (without beam). Taking into account all the factors mentioned above, we can express the relationship for the desired efficiency parameter ε as follows:

$$\varepsilon = \frac{N_{\Delta E}}{N_{tot} - \Delta N_1 - \Delta N_2} \qquad (6)$$

In this context, $N_{\Delta E}$ represents the number of signals with $\Delta E > 0$, and $\Delta N_{1,2}$ represents the values corresponding to the neutron-induced and escaping α-particle signals within the registered energy interval of 1.5-4.2 MeV. Meanwhile, $N_{tot}$ refers to the total number of events within that interval.

## 3. Summary

Various two-parameter function approximations have been tested to describe the background contribution in the region of registered α-decay peaks in the $^{232}$Th+$^{48}$Ca → Ds* complete fusion nuclear reaction. A comparison was made with the application of the VMRIA interactive peak searching program.

The presented algorithm allows for an express online estimation of the MWPC detection efficiency parameter. However, one partial disadvantage of the presented algorithm is that it operates well when the measured parameter of efficiency is close to or greater than ~0.7-0.8. In other words, the majority of signals within the considered energy interval should be correspond to charged particles flying from the cyclotron.

One extra conclusion can be drawn here. Namely, we plan to test the presented approach with function for background approximation differed from linear one in the variousheavy ions induced complete fusion nuclear reactions in a nearest future.A reasonable scenario is a power function with a power close to -0.5.Some additional tests with ΔE MWPC are in progress at DC-280 FLNR (JINR) cyclotron now, including more direct "insitu" measurements with heavy ions.

## 4. Supplement 1

An alternative scenario is to register target like products in ~25-45 MeV energy interval to estimate of an MWPC detection efficiency parameter. But, although this method is more direct and clear, one should take into account, that a time of several hours (up to ~ ten hours or even more) is require to obtain of the order of n·102 statistics under condition of ~3 pμA projectile intensity.